\documentclass{ptephy_v1}

\usepackage{latexsym,graphicx,amssymb,amsmath,mathrsfs}
\usepackage{epstopdf}
\usepackage{url}

\graphicspath{{./fig/}}

\newcommand\APF{\langle{e^{i\theta}}\rangle_\mathrm{pq}}
\newcommand\vev[1]{\langle #1 \rangle}
\newcommand\Pbar{\overline{P}}
\newcommand{\QCDone}{0+1D QCD}
\usepackage[normalem]{ulem}  

\renewcommand\sout{\bgroup \color{red} \ULdepth=-.5ex \ULset}

\begin{document}

\preprintnumber{KUNS-2761, YITP-19-34}

\title{
Path optimization in 0+1 dimensional QCD at finite density
}

\author[1]{Yuto Mori}
\affil{Department of Physics, Faculty of Science, Kyoto University,
Kyoto 606-8502, Japan
\email{mori.yuto.47z@st.kyoto-u.ac.jp}}

\author[2]{Kouji Kashiwa}
\affil{Fukuoka Institute of Technology, Wajiro,
Fukuoka 811-0295, Japan
\email{kashiwa@fit.ac.jp}}

\author[3]{Akira Ohnishi}
\affil{Yukawa Institute for Theoretical Physics, Kyoto University,
Kyoto 606-8502, Japan
\email{ohnishi@yukawa.kyoto-u.ac.jp}}

\begin{abstract}
We investigate the sign problem in 0+1 dimensional QCD
at finite chemical potential by using the path optimization method.
The SU(3) link variable is complexified to the SL(3,$\mathbb{C}$) link variable,
and the integral path is represented by a feedforward neural network.
The integral path is then optimized to weaken the sign problem.
The average phase factor is enhanced to be greater than 0.99
on the optimized path.
Results with and without diagonalized gauge fixing are compared and proven to be consistent.
This is the first step of applying the path optimization method to gauge theories.
\end{abstract}

\maketitle

\section{Introduction}
Solving the sign problem for complex actions in the path integral is
one of the grand challenges in theoretical physics.
When the Boltzmann weight is a complex-valued
highly oscillating function,
the strong cancellation takes place
and it is difficult to control the numerical error of integration.
The sign problem is a generic problem in
finite density fermion systems such
as finite density Quantum Chromodynamics (QCD).
There have been various approaches proposed so far
to avoid or evade the sign problem in finite density lattice QCD.
These approaches include
the reweighting method~\cite{Fodor:2001au,Fodor:2001pe,Fodor:2004nz,Fodor:2002km},
the Taylor expansion method~\cite{Miyamura:2002en,Allton:2005gk,Gavai:2008zr},
the analytic
continuation~\cite{deForcrand:2002ci,deForcrand:2003hx,D'Elia:2002gd,D'Elia:2004at,Chen:2004tb},
and
the canonical method by using the imaginary chemical
potential~\cite{Hasenfratz:1991ax,Alexandru:2005ix,Kratochvila:2006jx,deForcrand:2006ec,Li:2010qf,Nakamura:2015jra},
the density of states method~\cite{Gocksch:1988iz,Fodor:2007vv},
and the strong coupling QCD~\cite{Gagliardi:2017uag,Gagliardi:2018tkz,Ichihara:2014ova,Ichihara:2015kba,deForcrand:2014tha}.
Unfortunately, these methods work
only in the region of $\mu/T \lesssim 1$,
with heavy quark mass, or in the strong coupling regime
at present.

In addition, several approaches based on
complexified variables have been developed,
and recently attract much attention.
It may be possible to discuss physics
with severe sign problem.
These approaches include
the complex Langevin method~\cite{Parisi:1980ys,Parisi:1984cs},
the Lefschetz thimble method~\cite{Witten:2010cx,Cristoforetti:2012su,Fujii:2013sra},
and the path optimization method~\cite{Mori:2017pne,Mori:2017nwj,Kashiwa:2018vxr,Kashiwa:2019lkv}.
In this study, we use the path optimization method.

In the path optimization method,
the integration path (or manifold) is prepared appropriately,
and optimized so as to weaken the seriousness of
the sign problem.
The present authors have proposed this method~\cite{Mori:2017pne},
and have applied it to the one-dimensional integral
of a highly oscillating complex function~\cite{Mori:2017pne},
the two-dimensional scalar field theory at finite density~\cite{Mori:2017nwj},
and the Polyakov-loop extended Nambu-Jona-Lasinio
model~\cite{Kashiwa:2018vxr,Kashiwa:2019lkv}.
This method has been also applied to
the Thirring model~\cite{Alexandru:2018fqp} and
the one-dimensional complex scalar field theory
at finite density~\cite{Bursa:2018ykf}.

Now, it is important to discuss the sign problem in gauge theories
by using the path optimization method.
As the first step, we consider 0+1 dimensional QCD (\QCDone).
\QCDone\ has
the sign problem at finite chemical potential,
and the temporal hopping term of quarks is the origin of the sign problem,
as in 3+1 dimensional QCD.
The analytic solution is known for \QCDone~\cite{Bilic:1988rw},
then it is a good laboratory to examine
the framework for
the sign problem in gauge theory.
Actually, \QCDone\ has been studied 
by using 
the complex Langevin method~\cite{Aarts:2010gr},
the Lefschetz thimble method~\cite{DiRenzo:2017igr},
the subset method~\cite{Bloch:2013cxa,Bloch:2013ara},
and so on.

In this article,
as the first step towards investigating the sign problem
in gauge theories in $3+1$ dimension,
we apply the path optimization method to \QCDone\ at finite chemical potential.
We develop the method to complexify
the SU(3) link variable to that in SL(3,$\mathbb{C}$).
The integral path is prepared by using a feedforward neural network,
and is optimized to evade the sign problem.
We have found that the average phase factor on the optimized path
exceeds 0.99, which may allow us to study finite density
QCD at higher dimensions.
We also compare the results with and without diagonalized gauge fixing,
and the results are found to be consistent with each other.

This paper is organized as follows.
In Sec.~\ref{Sec:Formulation} and \ref{Sec:POM},
we explain \QCDone\ and the path optimization method, respectively.
We also discuss how to complexify the link variable.
In Sec.~\ref{Sec:Results},
we show the results with and without diagonalized gauge fixing,
and compare these results.
In Sec.~\ref{Sec:Summary},
we summarize this paper.

\section{$0+1$ dimensional QCD}
\label{Sec:Formulation}

In this study, we consider the $0+1$ dimensional QCD at finite
density with and without diagonalized gauge fixing in the path
optimization method.
Details are shown below.

\subsection{Partition function and observables}
The partition function of the \QCDone\ in the Euclidean spacetime
is given as~\cite{Bilic:1988rw,Faldt:1985ec},
\begin{align}
 {\cal Z}
 &= \int \mathcal{D}\overline{\chi}\, \mathcal{D}\chi\,
 \mathcal{D}U ~ e^{-S[\chi, \overline{\chi},U]},
 \label{eq:part_fn}
\end{align}
with the lattice action
\begin{align}
 S &= \frac{1}{2} \sum^{N_\tau}_{\tau=1}
      (\overline{\chi}_\tau e^{\mu} U_\tau \chi_{\tau+1} -
      \overline{\chi}_{\tau+1} e^{-\mu} U^{-1}_\tau \chi_\tau)
      + m \sum_\tau \overline{\chi}_\tau\chi_\tau,
 \label{eq:action}
\end{align}
where we consider one species of staggered fermion $\chi$,
$U_\tau$ is the $\mathrm{SU}(3)$ link variable in the temporal direction,
$m$ denotes the bare fermion mass,
$\mu$ is the chemical potential,
and $N_\tau$ is the number of lattice cites in the temporal direction.
By using the gauge transformation, one can reduce the degree of freedom
to one link variable,
and the partition function is rewritten as
\begin{align}
{\cal Z} 
=& \int dU\, \mathrm{det} D(U)
\ ,\label{Eq:Z}\\
D(U)
=& X + e^{\mu/T} U(\theta) + e^{-\mu /T}U^{-1}(\theta),
\end{align}
where $X = 2\cosh(E/T)$, $E = \mathrm{arcsinh}(m)$,
and $T = 1/N_\tau$ is the temperature.
The determinant can be written by using the Polyakov loop and its conjugate
for $N_c=3$,
\begin{align}
\det D(U)
=& X^3 + N_c X^2 (e^{\mu/T}P_U+e^{-\mu/T}\Pbar_U)
\nonumber\\
&+N_c X(e^{2\mu/T}\Pbar_U+e^{-2\mu/T}P_U+N_c\Pbar_U{P_U}-1)
\nonumber\\
&+2\cosh(N_c\mu/T)
+N_c e^{\mu/T}(N_c\Pbar_U^2-2P_U)
+N_c e^{-\mu/T}(N_cP_U^2-2\Pbar_U)
,\label{Eq:detD}
\end{align}
where $P_U=\mathrm{Tr}U/N_c$ and $\Pbar_U=\mathrm{Tr}U^{-1}/N_c$
are those before taking the average.
By using the one-link integral,
$\int dU U_{ab} U^{-1}_{cd}=\delta_{ad}\delta_{bc}/N_c$, 
we find $\int dU P=\int dU \Pbar=0$ and $\int dU \Pbar{P}=1/N_c^2$.
Then the integral of $\det D(U)$ is obtained as
$\mathcal{Z}=\int dU \det D(U)=X^3-2X+2\cosh(3\mu/T)$.

An analytic expression of the partition function is known as
\begin{align}
 {\cal Z} &= \frac{\sinh[(N_c+1)E/T]}{\sinh(E/T)} + 2\cosh(N_c\,\mu/T)
 ,
\end{align}
which agrees with the result for $N_c=3$
given in the previous paragraph.
Then
the quark condensate ($\sigma=\vev{\overline{\chi}\chi}$)
and the quark number density ($n_q$) in \QCDone\ are
obtained as
\begin{align}
 \sigma &=
 \frac{T}{V} \frac{\partial}{\partial m}\log Z,~~~
 n_q = - \frac{T}{V} \frac{\partial}{\partial \mu}\log Z,
\end{align}
where the spatial volume is unity in \QCDone, $V=1$.
The action in Eq.\,(\ref{eq:action}) is invariant under
the following transformation at $m=0$;
\begin{align}
 \chi_\tau &\to e^{i\epsilon_\tau \theta}\chi_\tau,~~~
 \overline{\chi}_\tau \to e^{i\epsilon\theta}\overline{\chi}_\tau,~~~
 \epsilon_\tau = (-1)^\tau.
\end{align}
The quark condensate $\langle\overline{\chi}\chi\rangle$ is
not invariant under this transformation, and can be regarded
as the order parameter of the symmetry breaking.
It should be noted that the above transformation is an analogue to
the chiral transformation in 3+1 dimensional QCD, by replacing
$\epsilon_\tau$ with $\epsilon_x = (-1)^{x_0+x_1+x_2+x_3}$.
Thus we refer to this as the chiral symmetry in \QCDone, and 
the quark condensate $\vev{\overline{\chi}\chi}$ is
referred to as the "chiral" condensate.
The other way to consider the chiral condensate in the
odd-dimensional system is using the higher dimensional
spinor-representation; see Ref.\,~\cite{Appelquist:1986fd} for
the $3$-dimensional system.

The Polyakov loop for $N_c=3$ is also obtained as
\begin{align}
P=
\Bigl\langle \frac{1}{N_c} \mathrm{Tr} \, U \Bigr\rangle
   = \frac{1}{N_c} \frac{(X^2-1)e^{-\mu/T}
     + X e^{2\mu/T}}{X^3-2X+2\cosh(3\mu/T)}\ .
\end{align}
which requires the technique developed 
in Ref.\, \cite{Bloch:2013ara,Nishida:2003fb}.  
It is also possible to obtain the Polyakov loop for $N_c=3$ by using
Eq.~\eqref{Eq:detD} and 
the one-link integral,
$\int dU U_{ab}U_{cd}U_{ef}=\varepsilon_{ace}\varepsilon_{bdf}/N_c!$,
which gives $\int dU P_U^3 = 1/N_c^3$.
It should be noted that there is no spatial dimensions,
and there is no "deconfinement" in \QCDone.
Thus we cannot regard the Polyakov loop as the order parameter
of the deconfinement. Nevertheless, the Polyakov loop represents
the energy of single quark excitation.

\subsection{Diagonalized gauge fixing}
By using the remaining gauge degrees of freedom,
the link variable can be diagonalized as
$U^\mathrm{diag}=\mathrm{diag}(e^{i\theta_1,},e^{i\theta_2},e^{i\theta_3})$,
then the integral in Eq.\,(\ref{eq:part_fn}) is rewritten as
\begin{align}
 \mathcal{Z}
 = \int d\theta_1 d \theta_2 \, H(\theta) \, e^{-S_\mathrm{eff}(\theta)},
\end{align}
where $S_\mathrm{eff}=-\log \mathrm{det}\,D(U^\mathrm{diag})$,
and $H(\theta)$ is the Haar measure given by
the Vandermonde's determinant~\cite{Kogut:1981ez},
\begin{align}
 H(\theta) = \prod_{a<b}^{3} \sin^2(\theta_a - \theta_b),~~
 \theta_3 = - \theta_1 - \theta_2.
\end{align}
The fermion determinant in this gauge is simply written as
\begin{align}
\det D(U^\mathrm{diag})=e^{-S_\mathrm{eff}}=\prod_{k=1}^{N_c}
(X+e^{\mu/T+i\theta_k}+e^{-\mu/T-i\theta_k})
\ .
\end{align}
In this diagonalized gauge, we have only two independent variables,
$\theta_1$ and $\theta_2$, and then explicit estimation of the integral
on mesh points is available.
We show the comparison between results
with and without the gauge fixing in the later discussions.

\section{Path optimization for 0+1 dimensional QCD}
\label{Sec:POM}
In the path optimization method, the trial functions, $z(x;C)$, which
represents the optimized integral path should be considered in the complex
integral-variable space where $C$ mean the parameters in the trial
function. Details are explained below.

\subsection{Cost function and reweighting}
The parameters $C$ are optimized to reduce the cost function
which represents the seriousness of the sign problem.
Specifically, we adopt the following cost function:
\begin{align}
 F_\mathrm{cost}(C)
 &= {\cal Z}_{pq}-|{\cal Z}|
 = \Bigl| \frac{{\cal Z}} {\langle e^{i \theta}\rangle_{pq}} \Bigr|
 - |{\cal Z}|,
\end{align}
with
\begin{align}
 {\cal Z}_{pq} &= \int dx\, \left| J(x;C) \, W(z(x;C)) \right|,\\
 e^{i\theta} &= \frac{J(x;C) \, W(z(x;C))}{|J(x;C) \, W(z(x;C))|},
\end{align}
where $W$ represents the Boltzmann weight and $\langle \cdots
\rangle_\mathrm{pq}$ denotes the phase quenched average;
\begin{align}
 \langle \mathcal{O} \rangle_{pq}
  &= \frac{1}{{\cal Z}_{pq}} \int dx \, \mathcal{O} \, |J(x;C) \, W(z(x;C))|.
\end{align}
Then, one can calculate the expectation value of observable as
\begin{align}
 \langle \mathcal{O} \rangle
  &= \frac{\int dx \, \mathcal{O}(x) \, W(x)}{\int dx \, W(x)}
  = \frac{\int dx \, J(x;C) \, \mathcal{O}(z(x;C)) \,W(z(x;C))}
          {\int dx \, J(x;C) \, W(z(x;C))}
  = \frac{\langle \mathcal{O}e^{i\theta} \rangle_{pq}}
          {\langle e^{i\theta} \rangle_{pq}},
\end{align}
By optimization of parameters, we can enhance the average phase factor,
and calculate the expectation values with small error, in principle.

\subsection{With diagonalized gauge fixing}
We shall now apply the path optimization method to \QCDone.
In the case with diagonalized gauge fixing,
we perform integration on the 2D mesh points,
\begin{align}
 {\cal Z} = (\delta x)^2 \sum_nJ(x^{(n)}+iy^{(n)})W(x^{(n)}+iy^{(n)}) .
\end{align}
where $n$ represents the mesh point. We give the imaginary parts $y_{1,2}$
by using the feedforward neural network,
\begin{align}
 y^{(n)}_i(x^{(n)};C)
 &= \omega_i f_i (x^{(n)}_1, x^{(n)}_2;C),
\ 
C=\left\{w, b, \omega\right\},
\end{align}
with $i=1,2$ and
\begin{align}
 f_i &= g(w^{(n,2)}_{ij} \, a^{(n)}_{j} + b^{(n,2)}_i), ~~~
 a^{(n)}_i  = g(w^{(n,1)}_{ij} \, x^{(n)}_{j} + b^{(n,1)}_i),
\label{Eq:FNN}
\end{align}
are the output variables of the input and hidden layers
in the feedforward neural network.
The variational parameters
$w$, $b$, $\omega$ are collectively denoted by $C$
and $g(\cdot)$ is called the activation function.
In this paper, we use the hyperbolic tangent function
as the activation function.
We include the Haar measure in the Boltzmann weight,
$W=He^{-S_\mathrm{eff}}$.
It is also possible to regard the imaginary parts $y^{(n)}$ as
the variational parameters when we fix the mesh points; see Methods
A and B explained in Sec.\,\ref{Sec:Results}.

\subsection{Without diagonalized gauge fixing}
In the case without diagonalized gauge fixing,
we parametrize the link variable as 
\begin{align}
 U \in \mathrm{SU(3)} \to \mathcal{U}(U) = U e^{y_1\lambda_1}
 \cdots e^{y_8\lambda_8} \in \mathrm{SL}(3,\mathbb{C}),
\label{Eq:U_SL3}
\end{align}
where $\lambda_a~(a = 1,...,8)$ are the Gell-Mann matrices and
the parameters $y_a$ are given by the feedforward neural network
in the same way as Eq.\,(\ref{Eq:FNN});
\begin{align}
 y_a = \omega_a
       f_a(\mathrm{Re}~U_{ij},\mathrm{Im}~U_{ij};C).
\end{align}

The Jacobian matrix can be calculated as follows.
If the local coordinate $x$ ($|x| \ll 1$) around
$U \in \mathrm{SU}(3)$ is given as $e^{ix_a\lambda_a}U$,
the Haar measure at $U$ is obtained as $d^8x=dx_1dx_2\cdots dx_8$.
Similarly, we consider the local coordinate $z$ around
$\mathcal{U}(U) \in \mathrm{SL}(3, \mathbb{C})$ as
$e^{iz_a\lambda_a}\mathcal{U}(U)$, where $z$ is given as a function of $x$
by the implicit equation,
$e^{iz_a\lambda_a}\mathcal{U}(U) = \mathcal{U}(e^{ix_a\lambda_a}U)$.
By differentiating this equation with
respect to $x$, we have
\begin{align}
 \frac{\partial z_b}{\partial x_a}\lambda_b e^{iz_c\lambda_c}\mathcal{U}(U)
 = \frac{\partial}{\partial x_a}\mathcal{U}(e^{ix_c\lambda_c}U).
\end{align}
The Jacobian matrix is then obtained as
\begin{align}
 J_{ab}
  &\equiv \left. \frac{\partial z_b}{\partial x_a} \right|_{x=0}
  = \frac{1}{2i} \mathrm{tr} \left. \left[ \left(
    \frac{\partial}{\partial x_a}\mathcal{U}(e^{ix_c\lambda_c}U)\right)
     \, \mathcal{U}^{-1}\lambda_b \right] \right|_{x=0}.
\end{align}
Therefore, the Haar measure at around
$\mathcal{U}(U)$ is given by $d^8z = \det (J)\,d^8x$.

To calculate the phase-quenched expectation value
and to evaluate the cost function,
we need the Monte-Carlo sampling,
\begin{align}
 \langle {\cal O} \rangle_{pq}
 = \frac{1}{N_\mathrm{conf}}\sum_k {\cal O}({\cal U}(U^{(k)}))
\end{align}
where $U^{(k)}$ denotes the $\mathrm{SU}(3)$ link variable in the $k$-th configuration
and the configurations $\{U^{(k)}\}$ are sampled according to
the probability distribution $|\det(J)\det(D(\mathcal{U}(U)))|$.
The details about the way to evaluate the cost function
$\mathcal{F}$ and to optimize the parameters $C$ stochastically
are explained in Ref.\ \cite{Mori:2017nwj}.

It should be noted that we complexify the $\mathrm{SU}(3)$ matrix $U$
to an $\mathrm{SL}(3;C)$ matrix $\mathcal{U}$, rather than complexifying $x_a$. 
For example, $y_a$ in Eq. \eqref{Eq:U_SL3}
is the imaginary part of $z_a$, complexified variable of  $x_a$,
only when $x$ and $y$ are small enough.
A more natural parametrization would be $\mathcal{U}(U)=\exp(iz_a \lambda_a)$
with $z_a=x_a+iy_a$, but the derivative of $\mathcal{U}$ with respect to $y$
becomes complicated in this parametrization.
Function forms other than that
in Eq.~\eqref{Eq:U_SL3} can be acceptable.
The parametrization dependence should be compensated by the Jacobian.

\section{Numerical results}
\label{Sec:Results}
As mentioned in the previous section,
we perform the numerical calculation in three different ways as follows.
\begin{description}
 \item[{\bf Method A}:]~\\
	The 2D mesh integral in the diagonalized gauge where
	the imaginary parts are regarded as the variational parameters.
 \item[{\bf Method B}:]~\\
	The 2D mesh integral in the diagonalized gauge where
	the imaginary parts are given by using the feedforward neural network.
 \item[{\bf Method C}:]~\\
	The Monte-Carlo sampling without diagonalized gauge fixing where
	the mapped SL(3) link variable is given by using the feedforward
	neural network. It should be noted
	that {\it this method is a promising way
		to be utilized
		in realistic $3+1$ dimensional QCD simulation}.
\end{description}
We first summarize our numerical setup below,
and next we show our numerical results.

\subsection{Numerical setup}
In the Methods A and B, we utilize the standard gradient descent
method to optimize the imaginary parts via the equation,
$\dot{C} =-\partial F_\mathrm{cost}/\partial C$,
which shows the fictitious time evolution.
In the Method A, we prepare $30\times30$ or $60\times60$
mesh points to perform the integration,
and we update the parameters
with the fictitious time step
$\Delta t = 10^{-3} \sim 10^{-2}$.
We also invoke smearing of the imaginary part.
When the average phase factor is larger after averaging the imaginary part
with the adjacent mesh points, we adopt that configurations
as a further optimized path.
In the Method B, the number of mesh points is $25\times25$,
and the learning rate is set to $\Delta t = 10^{-3} \sim 10^{-2}$.
In the Method C, the optimization is performed by using the stochastic
gradient descent method.
Actually, we employ Adadelta~\cite{zeiler2012adadelta} as the optimizer.
Parameters
in the \QCDone\ action
are given as $m = 0.05, T = 0.5\,(N_\tau=2)$,
and the chemical potential range of $\mu/T = 0\sim 2$ is discussed.
We note that the quark number density is almost saturated
in the region $\mu/T > 2$.

\subsection{Results with diagonalized gauge fixing: Methods A and B}

\begin{figure}[bthp]
\begin{center}
\includegraphics[width=0.33\textwidth]{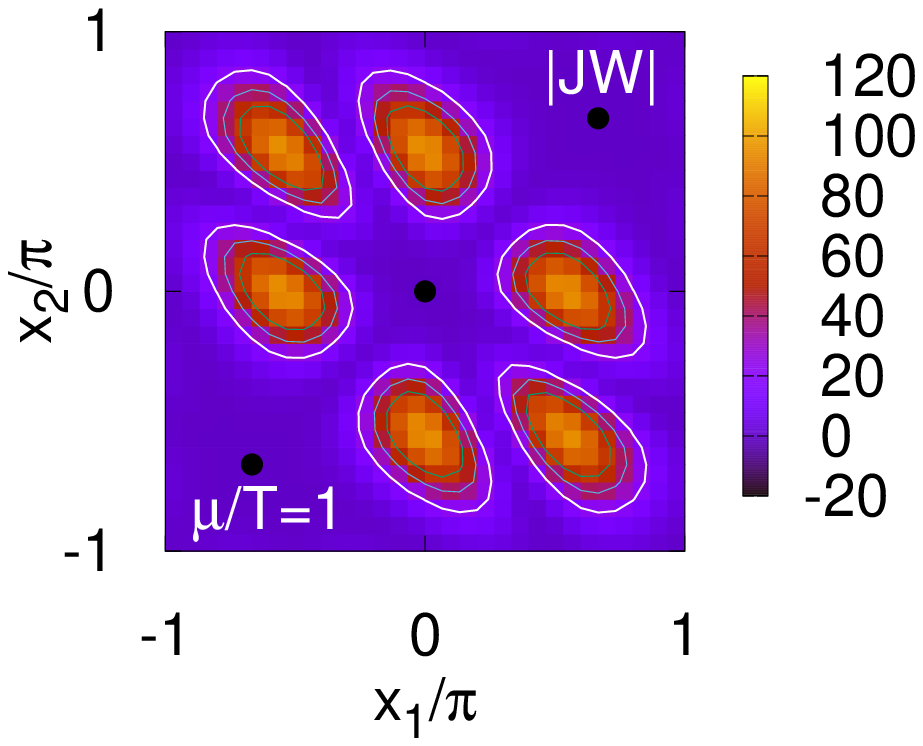}%
\includegraphics[width=0.33\textwidth]{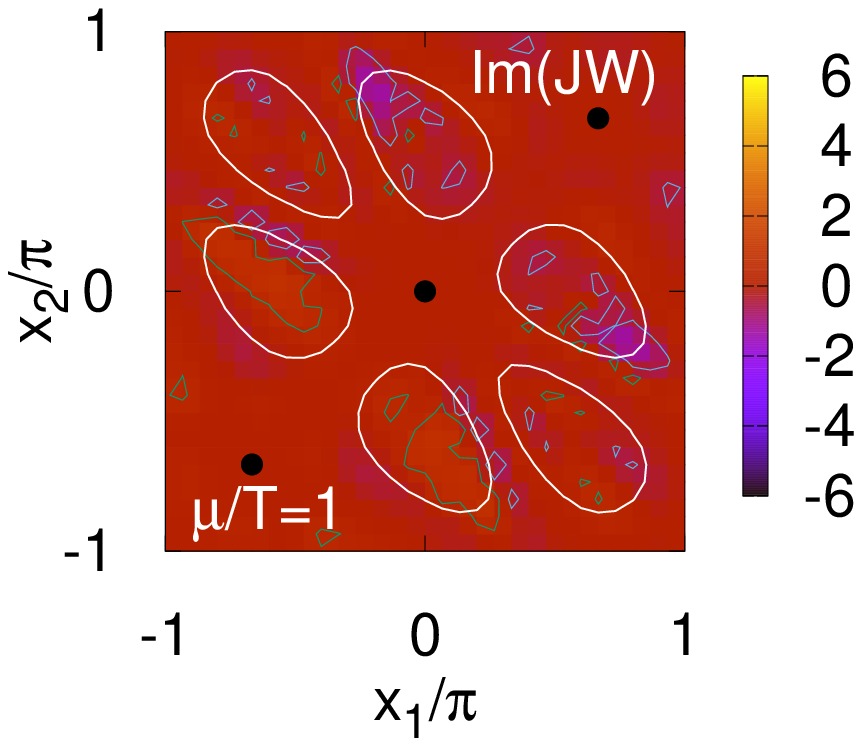}%
\includegraphics[width=0.33\textwidth]{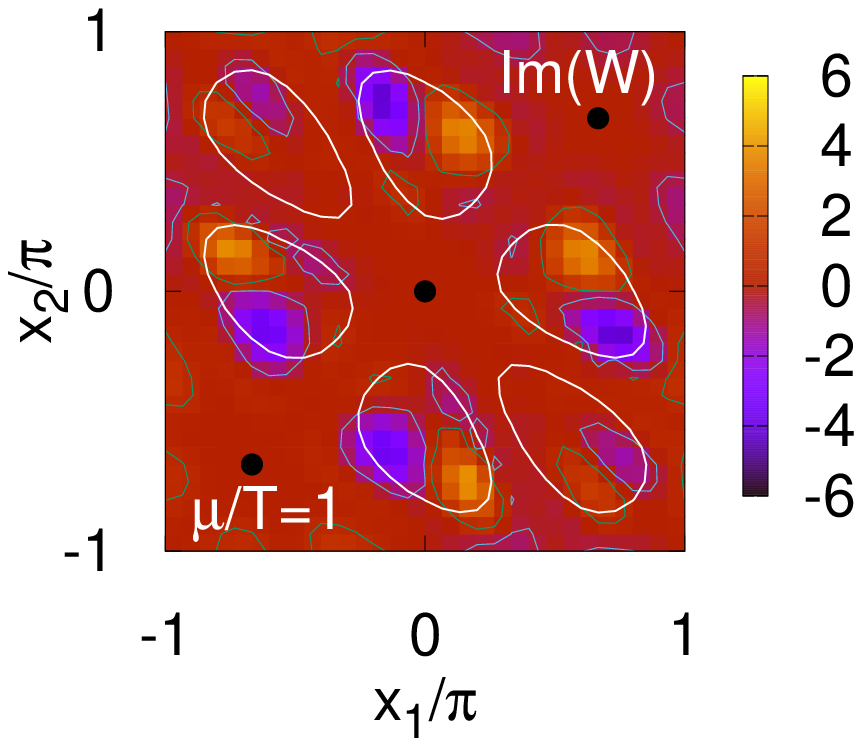}%
\end{center}
\caption{
Absolute value of the statistical weight, $|JW|$ (left),
imaginary part of the statistical weight, $\mathrm{Im}(JW)$ (middle),
and imaginary part of the Boltzmann weight, $\mathrm{Im}(W)$ (right),
on the optimized path.
We show the results at $\mu/T=1$ on a $30^2$ lattice.
The optimized path is obtained in the Method A.
White curves show the contour of $|JW|=20$.}
\label{Fig:2D_JW}
\end{figure}

Let us now discuss the numerical results in the diagonalized gauge.
In Fig.~\ref{Fig:2D_JW},
we show the absolute value and the imaginary part of the statistical weight $JW$
on the optimized path.
In the diagonalized gauge,
there are six separated regions on the $(x_1,x_2)$ plane
where $|JW|$ is significantly large as shown in the left panel.
Compared with the absolute value, the imaginary part is suppressed strongly
as shown in the middle panel.
This suppression comes mainly from the suppression of the complex phase
of the Boltzmann weight, and in part from the cancellation of the complex phase
of the Jacobian and the Boltzmann weight.
In the right panel of Fig.~\ref{Fig:2D_JW},
we show the imaginary part of the Boltzmann weight, $\mathrm{Im}(W)$,
which has larger absolute values compared with $\mathrm{Im}(JW)$.
This is one of the merits of using the path optimization method:
The optimization including the complex Jacobian effects
leads to smaller imaginary part of $JW$.

\begin{figure}[htbp]
\begin{center}
\includegraphics[width=0.45\textwidth]{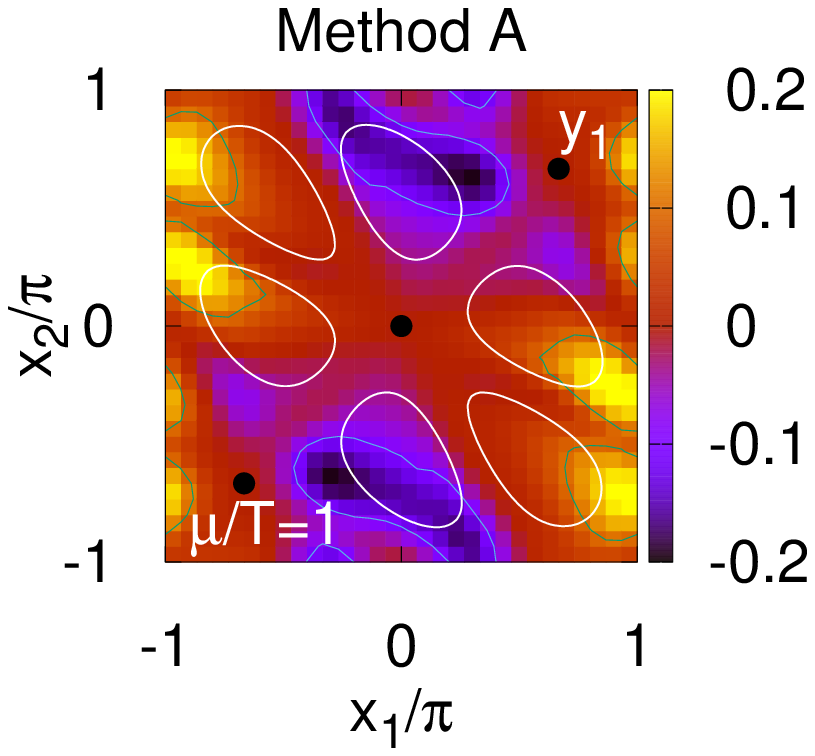}
\includegraphics[width=0.45\textwidth]{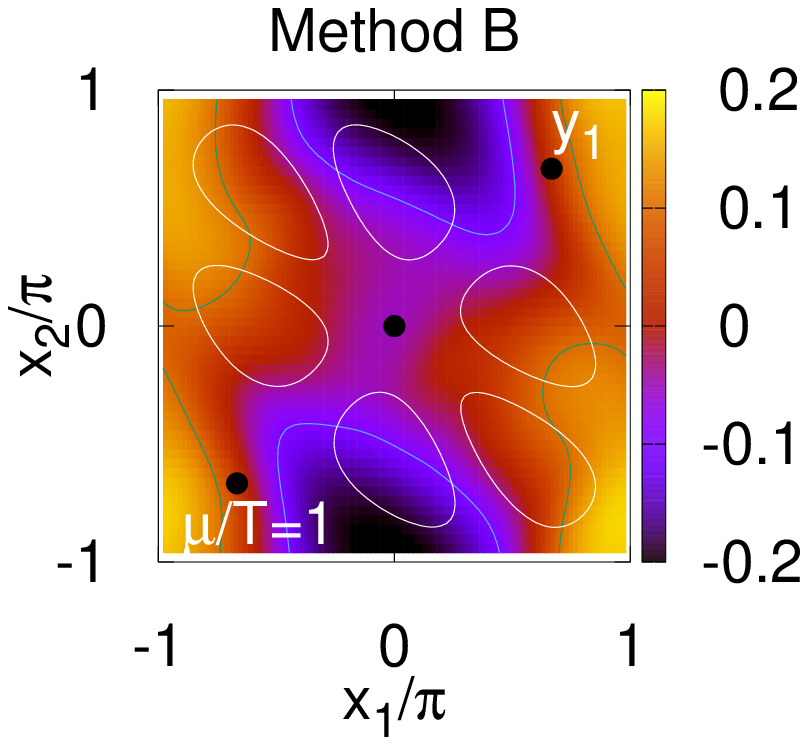}
\end{center}
\caption{
Imaginary part of $z_1$ on the optimized path.
We show $y_1$, the imaginary part of $z_1$, on the optimized path
as a function of the real parts
on a $30^2$ lattice at $\mu/T=1$ in the Methods
A (left) and B (right).
White curves show the contour of $|JW|=20$.
}
\label{Fig:2D_y1}
\end{figure}

In Fig.\,\ref{Fig:2D_y1},
we show $y_1$, the imaginary part of $z_1$,
as a function of real parts, $(x_1,x_2)$.
The imaginary part of $z_2$ is obtained by exchanging $x_1$ and $x_2$.
We compare the results in the Method A (left)
and those in the Method B (right).
The standard gradient descent method and the feedforward neural network
give qualitatively the same but somewhat different results of $y_1$.
In the statistically significant region (inside the white curve),
these results are more consistent with each other.
Since the cost function does not have sensitivity to the region
with small weight, results can be different outside of the statistically
significant regions as we expected.

\begin{figure}[htbp]
\begin{center}
\includegraphics[width=0.45\textwidth]{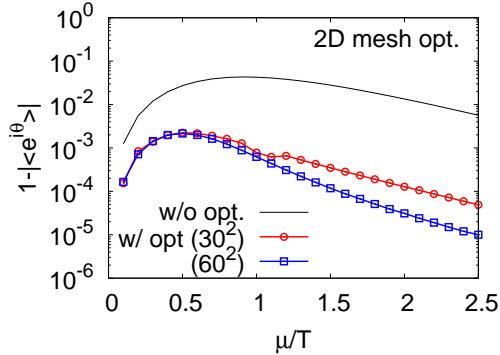}
\end{center}
\caption{
Average phase factor after path optimization
on 2D mesh points in the Method A.
Results with mesh points of $30^2$ (red circles) and $60^2$ (blue squares)
are shown in comparison with the average phase factor without optimization.
We show the difference of the absolute values of the average phase
factor from unity, $1-|\vev{e^{i\theta}}|$.}
\label{Fig:2Dmesh}
\end{figure}

In Fig.~\ref{Fig:2Dmesh},
we show the average phase factor as a function of $\mu/T$
obtained on the $30^2$ and $60^2$ mesh points in the Method A.
We show the difference from unity of the average phase factor.
After optimization, the average phase factor is enhanced
to be greater than $0.997$, i.e. $1-|\APF| < 3\times 10^{-3}$.
While the average phase factor is large even without optimization,
$|\APF|>0.95$, the difference with and without optimization appears
strongly in finite spatial volume cases, as discussed later.
The minimum average phase factor appears at around $\mu/T=0.5$.

As already mentioned,
we clearly see that
the statistical weight has six separated regions
on the $(x_1,x_2)$ plane in the diagonalized gauge.
This separation comes from the structure of the Haar measure
and it causes the numerical problem in Monte-Carlo sampling:
it is difficult to sample all of relevant regions beyond
the energy barriers between them.
It is, of course, not a problem
in the mesh point integration,
but we cannot apply the mesh point integration
to more realistic problems in field theories because
of its enormous numerical cost. Thus we proceed to discuss
the results based on the Monte-Carlo integral without the diagonalized
gauge fixing.

\subsection{Results without diagonalized gauge fixing: Method C}

\begin{figure}[htbp]
\begin{center}
\includegraphics[width=0.45\textwidth]{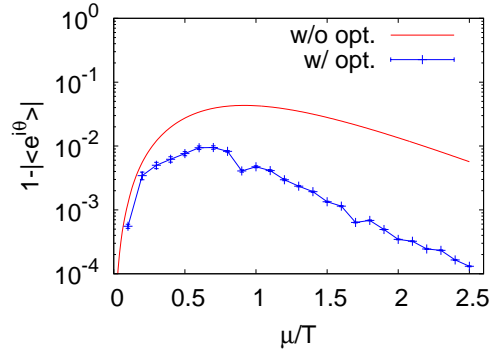}
\end{center}
\caption{
The average phase factor with (blue dots)
and without (red line) optimization.
Results with optimization are obtained in the Method C.
The expectation value without the optimization is
obtained by using the 2D mesh point integral
with diagonalized gauge fixing.}
\label{Fig:apf}
\end{figure}

Next, we discuss the results without diagonalized gauge fixing.
Since we have eight independent variables,
the mesh point integration is not possible to perform.
We utilize the feedforward neural network to prepare and optimize 
the path, and we apply the hybrid Monte-Carlo method
to sample the configurations in the Method C.
Figure~\ref{Fig:apf} shows the average phase factor with and without
the path optimization.
The path optimization is performed by using the method C and the results
without the path optimization is estimated by using the 2D
mesh point integration.
We can clearly see that the path optimization increases the average phase
factor in all values of $\mu/T$ shown in the figure.

\begin{figure}[htbp]
\begin{center}
\includegraphics[width=0.50\textwidth]{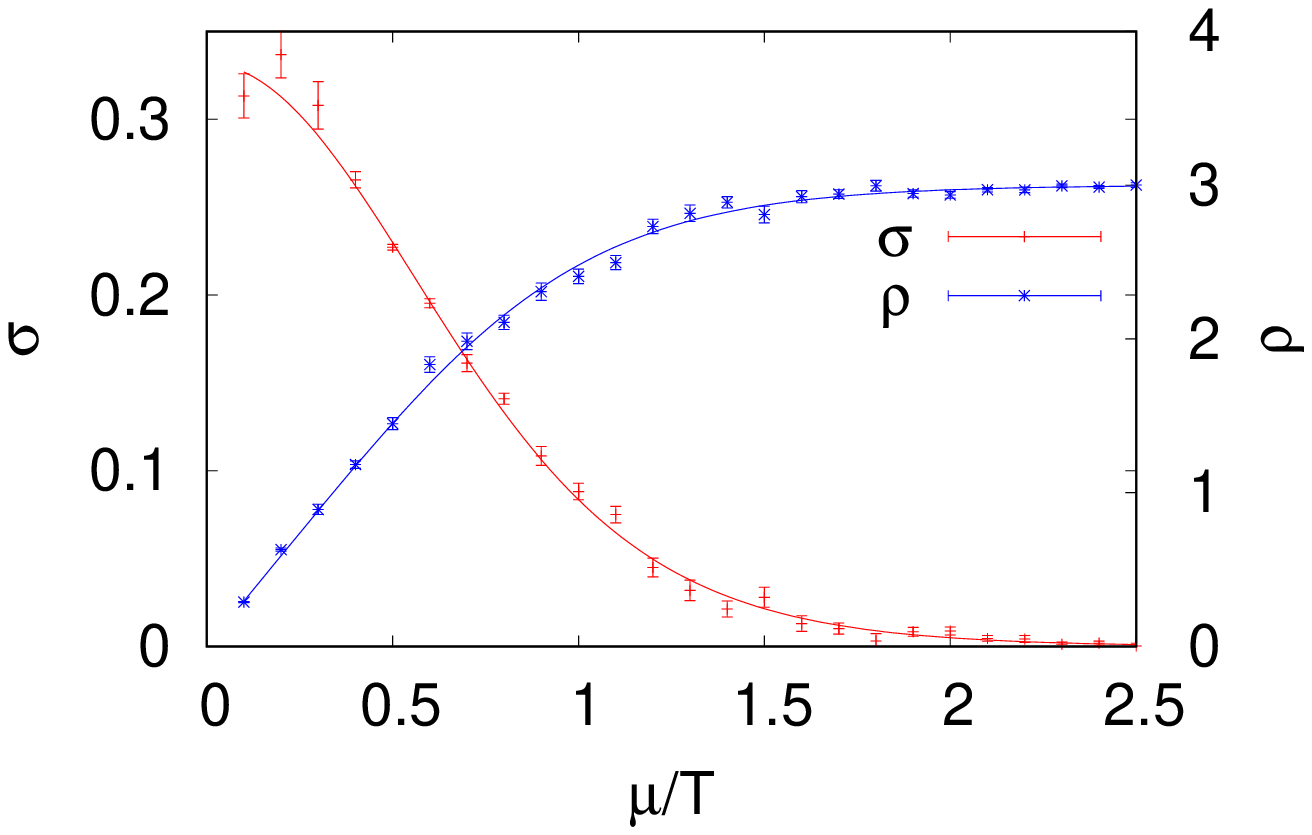}
\includegraphics[width=0.45\textwidth]{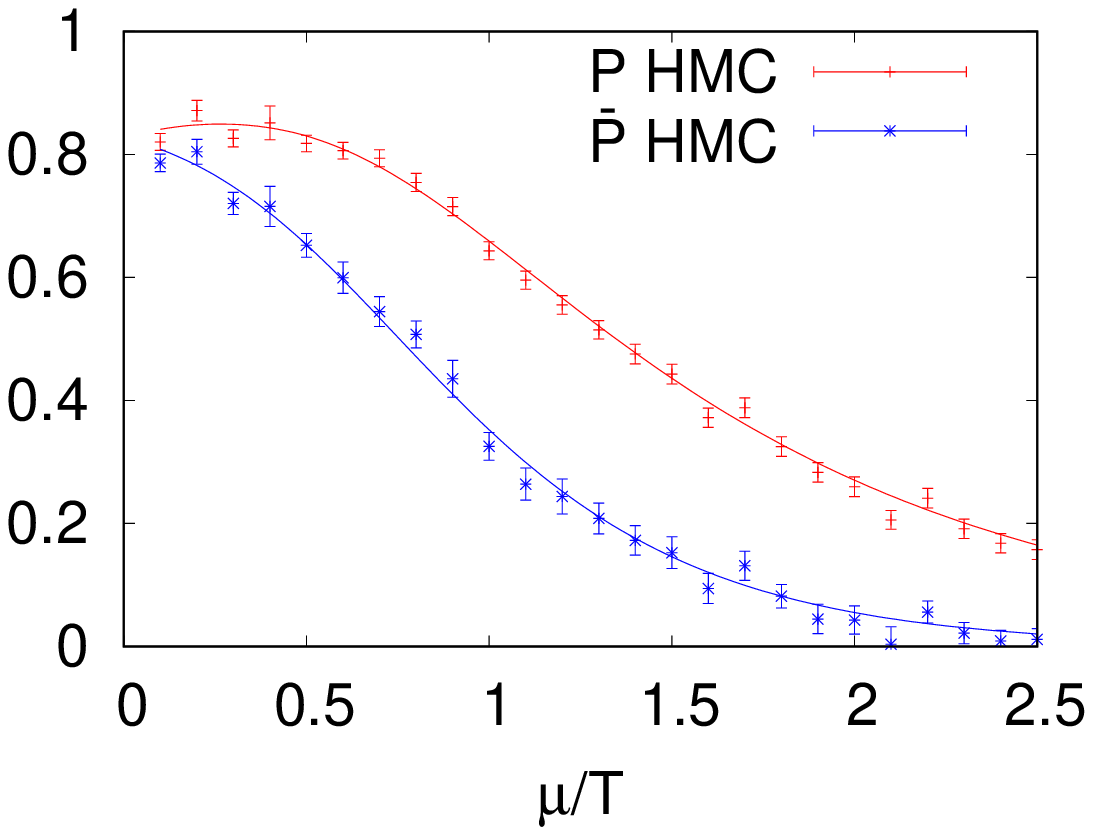}
\end{center}
\caption{
The expectation values of the quark condensate,
the quark number density, and the Polyakov loop.
The symbols show numerical results by using the hybrid Monte-Carlo method,
and the lines are analytic exact results.
}
\label{Fig:expect}
\end{figure}

In Fig.\,\ref{Fig:expect}, we show the expectation values of
the chiral condensate ($\sigma$),
the quark number density ($\rho$),
and the Polyakov loop ($P$),
as functions of $\mu/T$
in comparison with exact results.
Here, we use $N_\mathrm{conf}=1000$ for $\sigma$ and $\rho$ and
$N_\mathrm{conf}=10000$ for $P$.
Since the fluctuation of $P$ is larger than those
of $\sigma$ and $\rho$, we need larger number of configurations.
The expectation values on the optimized integral path
agree well with exact results within the error bar.
The chiral condensate decreases rapidly
in the $\mu/T =0.5\sim1.0$ region
and the Polyakov loop seems to
stay at small values above $\mu/T > 2$.

\begin{figure}[htbp]
\begin{center}
\includegraphics[width=0.50\textwidth]{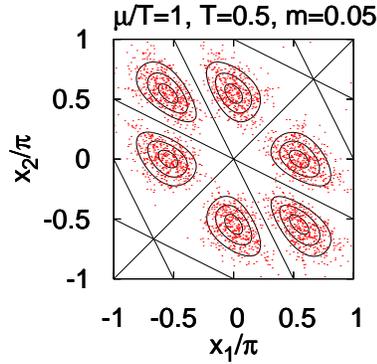}
\end{center}
\caption{
The scatter plot of $(x_1,x_2)$ in the Method C.
The variables $(x_1,x_2)$ are obtained
by diagonalizing the link variables,
specifying the eigenvalues as $(e^{iz_1}, e^{iz_2}, e^{iz_3})$,
and taking the real parts of $(z_1,z_2)$.
The configurations are generated by using the hybrid Monte-Carlo method.
}
\label{Fig:dist_wo_gauge}
\end{figure}

It is interesting to check if
we can reproduce the statistical weight distribution
obtained in the diagonalized gauge (Methods A and B) shown in
the left panel of Fig.~\ref{Fig:2D_JW}, by using
configurations obtained in the method C.
We diagonalize the link variable by the similarity transformation,
$\mathcal{U}=P\mathcal{U}^{\mathrm{diag}}P^{-1}$,
and $(z_1,z_2)$ are specified by the diagonal matrix elements,
$\mathcal{U}^\mathrm{diag}=\mathrm{diag}(e^{iz_1},e^{iz_2},e^{iz_3})$.
Figure\,\ref{Fig:dist_wo_gauge} shows the scatter plot
of the real part of $z_1$ and $z_2$.
From the figure, we can see that the relevant six regions are
actually visited in the Markov-chain hybrid Monte-Carlo sampling.

\section{Summary}
\label{Sec:Summary}

In this article,
we have examined the validity and usefulness of the path optimization
method in the $0+1$ dimensional QCD.
The $0+1$ dimensional QCD is a toy model of
the realistic $3+1$ dimensional QCD, but it contains the same terms,
the temporal hopping terms of quarks,
which causes the
sign problem in $3+1$ dimension;
we can use it as a good laboratory to investigate several
issues of the sign problem.

We have found that the path optimization method combined with
the feedforward neural network
is a useful tool to study the sign problem from several reasons.
First,
it can well improve the average phase factor.
Second, the statistical weight distributions with and without
diagonalized gauge fixing agree with each other.
We find that,
in \QCDone\ with diagonalized gauge fixing,
the statistical weight distribution is separated into six regions,
which could prevent us from sampling configurations equally well
from these regions in Markov-chain Monte-Carlo methods.
However, the six regions
are found to be well visited in Monte-Carlo configurations
generated without diagonalized gauge fixing.
This fact indicates
that the energy barriers
in the variable space with the diagonalized gauge, $(x_1,x_2)$,
are apparent ones, and do not cause trouble in gauge unfixed calculations.
Thirdly, some of the observables are demonstrated to be obtained correctly.
We have calculated the chiral condensate, the quark number density,
the Polyakov loop and its conjugate as functions of $\mu/T$,
and these results reproduce the exact ones within the errors.
This is not surprising, since the integral of holomorphic functions are
independent of the path as long as the path does not go through the singular
points of the Boltzmann weight, which do not exist in the region
of finite imaginary parts of integral variables.

The average phase factor without the gauge fixing on the optimized path
is found to be larger than 0.995.
This value is corresponding to the situation that
the average phase factor is about $\sim 0.08$ on the $8^3\times N_\tau$
lattice, provided that other terms in the 3+1 dimensional QCD do not
make the average phase factor worse.
The average phase factor of $0.08$
seems to be small, but it is not impossible to perform the
Monte-Carlo sampling in the current computer power.
Compared with the 2D mesh point integration,
the average phase factor is still smaller.
Then it would be possible to further optimize the integral path
by, for example, modifying the cost function to be more sensitive
to the average phase factor at around $\APF \simeq 1$.
The path optimization method, thus, may be a promising method
to overcome the sign problem in realistic QCD,
if we can reduce the numerical
cost; for example, we need to reduce the Jacobian computation.
The promising way is
the diagonal ansatz of the Jacobian matrix~\cite{Alexandru:2018fqp}
and the nearest-neighbor lattice-cite ansatz~\cite{Bursa:2018ykf}.

\section*{Acknowledgments}
This work is supported in part by the Grants-in-Aid for Scientific Research
 from JSPS (Nos. 
15H03663, 
16K05350, 
18J21251, 
18K03618, 
and
19H01898), 
 and by the Yukawa International Program for Quark-hadron
 Sciences (YIPQS).

\bibliographystyle{ptephy}
\bibliography{ref}

\end{document}